**Title:**
Genome-wide signals of pervasive positive selection in human evolution

**Authors:**
David Enard*, Philipp W. Messer, and Dmitri A. Petrov

**Affiliations:**
Department of Biology, Stanford University, Stanford, CA 94305, USA
*To whom correspondence should be addressed at denard@stanford.edu

**Abstract:**

The role of positive selection in human evolution remains highly controversial (Cai et al. 2009; Hernandez et al. 2011; Lohmueller et al. 2011). On the one hand, scans for positive selection have identified hundreds of candidate loci and the genome-wide patterns of polymorphism show signatures consistent with frequent positive selection. On the other hand, recent studies have argued that many of the candidate loci are false positives and that most apparent genome-wide signatures of adaptation are in fact due to reduction of neutral diversity by linked recurrent deleterious mutations, known as background selection (Hernandez et al. 2011). Here we analyze human polymorphism data from the 1,000 Genomes project (Abecasis et al. 2012) and detect signatures of pervasive positive selection once we correct for the effects of background selection. We show that levels of neutral polymorphism are lower near amino acid substitutions, with the strongest reduction observed specifically near functionally consequential amino acid substitutions. Furthermore, amino acid substitutions are associated with signatures of recent adaptation that should not be generated by background selection, such as the presence of unusually long and frequent haplotypes (measured with *iHS* and *XPEHH*) and specific distortions in the site frequency spectrum (measured with *CLR*). We use forward simulations to show that the observed signatures require a high rate of strongly adaptive substitutions in the vicinity of the amino acid changes. We further demonstrate that the observed signatures of positive selection correlate more strongly with the presence of regulatory sequences, as predicted by ENCODE (Gerstein et al. 2012), than the positions of amino acid substitutions. Our results establish that adaptation was frequent in human evolution and provide support for the hypothesis of King and Wilson (King and Wilson 1975) that adaptive divergence is primarily driven by regulatory changes.

**Introduction.**

The rate and patterns of positive selection are of fundamental interest for the study of human evolution. Population genomic studies should in principle allow us to quantify positive selection from its expected signatures in sequence polymorphism and divergence data. Surprisingly, despite sequencing of thousands of human genomes (Abecasis et al. 2012) and the availability of whole genome sequences of closely related species, the extent to which adaptation has left identifiable signatures in the patterns of polymorphism in the human genome remains highly controversial (Akey 2009; Hernandez et al. 2011).

On the one hand, recent studies have identified a large number of loci showing signatures of recent selective sweeps (Voight et al. 2006; Sabeti et al. 2007; Williamson et al. 2007; Pickrell et al. 2009; Grossman et al. 2013) and McDonald-Kreitman (MK) analyses inferred that ~10-20% of amino acid changes have been adaptive in human evolution (Boyko et al. 2008). Consistently, regions of high functional density, high rate of amino acid substitutions, and low recombination all show reduced levels of neutral diversity (Cai et al. 2009; Lohmueller et al. 2011), as expected under recurrent selective sweeps in functional regions.

On the other hand, there are reasons to question that adaptation left clear signatures in the human genome. First, different scans for positive selection have identified largely non-overlapping sets of candidates (Akey 2009), which could be due to a high rate of false positives. Second, MK analyses can be confounded by a number of factors, such as perturbations left by demographic events and by the presence of slightly deleterious mutations (Eyre-Walker and Keightley 2009; Messer and Petrov 2013), and some MK analyses have failed to find evidence for adaptation in the human lineage (Eyre-Walker and Keightley 2009). Finally, it has been argued that background selection (BGS) (Charlesworth et al. 1993), a process in which deleterious mutations remove linked neutral variation from the population, should reduce levels of polymorphism in regions of higher functional density and low recombination, providing an alternative explanation for the observation of these correlations in the human genome.

One signature of positive selection - lower levels of neutral variation near functional substitutions (Andolfatto 2007; Macpherson et al. 2007; Cai et al. 2009) - is not generally expected under BGS and should therefore provide the clearest genomic evidence for the action of positive selection. While this signature was found in the human genome by Cai et al (Cai et al. 2009), it could not be detected by two recent studies using the newest large-scale datasets of human diversity (Hernandez et al. 2011; Lohmueller et al. 2011). In particular, Hernandez *et al.* (Hernandez et al. 2011) searched for lower levels of neutral diversity near functional substitutions by contrasting levels of neutral diversity near nonsynonymous compared to synonymous substitutions, following the study design of Sattath *et al.*(Sattath et al. 2011). They failed to find this signature in the human genome, and, moreover, found that diversity might in fact be marginally higher near nonsynonymous substitutions. Simulations showed that this puts sharp limits on the amount of adaptation by classic selective sweeps in recent human evolution (Hernandez et al. 2011).

However, it is likely that the study design of Sattah *et al.* (Sattath et al. 2011) is strongly biased against finding signatures of positive selection in the human genome and all other genomes with sharply variable levels of genomic constraint. This is because, as we show in the Results, nonsynonymous substitutions in the human genome tend to be located in regions of weaker constraint and thus weaker BGS compared to synonymous substitutions. These differences in levels of BGS should elevate neutral diversity near nonsynonymous compared to that near synonymous substitutions. The approach of Sattath *et al.* would thus detect positive selection only if the reduction of diversity due to positive selection near nonsynonymous substitutions happens to be greater than the initial difference in the opposite direction due to BGS.

Here we employ a number of more sensitive approaches in the search for signatures of positive selection while attempting to reduce the confounding effects of BGS to the greatest extent possible. Our results suggest that positive selection was frequent in human history and likely involved adaptive mutations of substantial selective effect. We estimate that on the order of a few hundred of strong adaptive events are likely to be detectable in the human genome, consistent with the latest scan for positive selection (Grossman et al. 2013). Moreover, we show that the majority of adaptive substitutions likely resulted in cis-regulatory rather than protein-coding changes, providing evidence in favor of the King and Wilson (King and Wilson 1975) hypothesis that adaptive divergence is primarily driven by regulatory changes.

**Results.**

The search for signals of positive selection in the human genome is complicated by the high variability of functional constraint and thus highly variable and tightly correlated levels of BGS and genetic draft across the genome. BGS should correlate most strongly with the density of constrained elements, while the strength of genetic draft (proportional to the product of the rate and strength of selective sweeps) should correlate with the number of functional substitutions. However, in the human genome the correlation coefficient between the amount of coding sequences (CDS) per 400Kb window and the number of nonsynonymous substitutions is ~0.89 (Cai et al. 2009), making it very difficult to separate effects of BGS and genetic draft from each other.

Moreover, the prediction of lower neutral polymorphism near nonsynonymous substitutions (Andolfatto 2007; Macpherson et al. 2007; Cai et al. 2009) is biased against detection of positive selection in the presence of BGS. This is because, even under the model of frequent positive selection, many or even most nonsynonymous substitutions are likely to be neutral in human evolution (Boyko et al. 2008). These neutral substitutions, in turn, should preferentially mark regions of weaker functional constraint, and therefore weaker BGS and higher levels of polymorphism. Thus, the reduction of neutral polymorphism near nonsynonymous substitutions needs to be stronger than the elevation of polymorphism due to weaker BGS in the same regions.

This bias against detecting evidence of genetic draft should be particularly strong when levels of neutral polymorphism near synonymous and nonsynonymous substitutions are contrasted, as in the approach pioneered by Sattath *et al.* (Sattath et al. 2011) and applied by Hernandez *et al*. (Hernandez et al. 2011) to the human 1,000 genomes pilot data. First, in the human genome the majority (~65%) of all synonymous substitutions are located extremely close (less than 0.02 cM) to nonsynonymous ones (Methods), thereby substantially reducing the power of this method. What is more troubling is that the synonymous substitutions that are located far from nonsynonymous ones, which in principle could provide a reasonable statistical control, mark regions of particularly strong selective constraint and thus particularly strong BGS. Specifically, we find that 82% of synonymous substitutions are found within conserved segments of the genome predicted by phastCons (Siepel et al. 2005) (Methods), versus only 56% of nonsynonymous ones.

The reason for this difference between synonymous and nonsynonymous substitutions is that selectively constrained regions, by definition, lack nonsynonymous substitutions, but still allow changes at unconstrained synonymous sites. Because constrained regions have stronger BGS, this means that in the absence of positive selection, heterogeneity in BGS should reduce diversity near synonymous substitutions to a greater extent compared to diversity near nonsynonymous substitutions. This pattern is in fact observed by Hernandez *et al.* (Hernandez et al. 2011) and suggests that use of the synonymous control in the functionally heterogeneous human genome might be unduly conservative.

Below we devise more sensitive methods for the detection of positive selection in the human genome. We first search for reduction of neutral polymorphism near all, and then also near functionally important, nonsynonymous substitutions. Specifically, we match regions near and far from nonsynonymous substitutions by levels of BGS measured using a variety of correlates of BGS, such as levels of functional constraint and recombination. We focus specifically on regions of low BGS, because in these regions the bias against finding positive selection should be the weakest. We then estimate haplotype statistics *iHS* and *XPEHH*. Unlike the overall level of neutral polymorphism employed in the first set of tests, we demonstrate that these haplotype statistics are virtually insensitive to BGS and, as a result, that their extreme deviations are strongly predictive of recent and strong selective sweeps. Finally, we quantify the rate and strength of positive selection required to produce the extent of the signatures of genetic draft we detect.

All data analyses are carried out with the 1,000 Genomes phase 1 data 20100804 release (http://www.1000genomes.org). Levels of neutral diversity are calculated as the average pairwise heterozygosity at putatively neutral sites scaled by divergence between human and macaque. Human specific substitutions at synonymous and nonsynonymous sites are inferred from human-chimpanzee-orangutan alignments (Methods).

*Choosing analysis windows*
BGS is expected to be stronger in regions of low recombination: consistently with this, the correlation between neutral diversity and recombination rate measured in 500 kb windows sliding every 50kb is strong and positive ($n$=44,958, Spearman's $\rho$=0.43, $P$<2x10$^{-16}$). BGS should also be stronger in regions of high functional constraint. We can measure functional constraint using multiple variables, all of which show strong negative correlations with levels of neutral diversity in 500 kb windows, including (i) density of coding sequences (CDS) ($n$= 44,958, $\rho$=-0.22, $P$<2x10$^{-16}$), (ii) density of conserved coding and non-coding sequences (CCDS) in all mammals or just in primates according to phastCons ($\rho$=-0.22, $P$<2x10$^{-16}$ and $\rho$=-0.26, $P$<2x10$^{-16}$, respectively), and (iii) the density of UTRs ($\rho$=-0.22, $P<2\times10^{-16}$). All of these correlations were computed as partial correlations controlling for recombination rate. In addition, we also find a strong negative partial correlation between diversity and GC content controlling for recombination ($\rho$=-0.19, $P<2\times10^{-16}$) that might be related to high GC content of coding regions (Lander et al. 2001) or some other property correlated with the GC content.

The segments of conserved DNA identified by phastCons are shared by mammals and/or primates and represent averaged constraint over long evolutionary periods of time. The density of mammalian and primate constrained sequences being equal, regions that are particularly devoid of human-specific nonsynonymous substitutions may be under stronger recent constraint. To detect such regions of unusually strong constraint, and thus BGS, we plot the distribution of distances to the nearest amino acid substitution in the human genome in all regions that have CCDS density greater than 0.1% (to make sure there are CCDS in the windows). 67.4% of the windows are located less than 0.1 cM away from a human-specific amino acid substitution, 30% are located 0.1 to 1 cM, and 2.6% are located further than 1 cM. These latter windows may represent regions of unusually strong conservation in the human lineage.

We quantify whether the regions of moderate to high functional density (CCDS density>0.5%) located far (> 1 cM) from any amino acid substitution are indeed subject to stronger BGS by conducting a bootstrap procedure (Methods). For each window located between 0.1 and 1 cM away from an amino acid change we match a randomly sampled window located 1 cM or further whose functional density, GC content, and recombination do not differ by more than empirically fixed thresholds compared to the 0.1cM - 1 cM window. Windows less than 0.1 cM away are excluded from this comparison since they are the ones most likely to be affected by positive selection and we want to focus only on BGS as a function of distance to the nearest amino acid change. Thresholds of the bootstrap are adjusted such that 0.1 cM to 1 cM and >1 cM windows have similar average functional density, GC content and recombination rates. Windows for which no good match can be found are excluded (the detailed bootstrap procedure is described in the Methods).

Neutral diversity is indeed substantially reduced in regions that are located more than 1 cM away from any amino acid substitution, controlling for functional density, GC content, and recombination. Overall, the reduction is 7% (randomization test, $P=8.6\times10^{-3}$) and becomes even stronger (~15%; randomization test, $P=2.4\times10^{-2}$) in regions where recombination rates do not exceed 1 cM/Mb. This is consistent with our interpretation that regions of substantial functional density that are located very far from any amino acid change are more constrained in the human lineage in a way that cannot be accounted for using levels of constraint in more distant mammalian and primate species. Below we exclude the 2.6% of the windows that are located further than 1 cM from an amino acid substitution.

*The near-vs-far test.*

The key expectation of positive selection is that it should reduce neutral diversity near functional substitutions. We first test this prediction by contrasting neutral diversity in 500kb windows near (< 0.1 cM) compared to far (> 0.5 cM) from any of the 21,278 amino-acid substitutions we identified in the human lineage (Methods). Importantly, the windows are matched by all parameters associated with BGS that we described above.

We first carry out this test in the regions with low density of conserved coding sequences (CCDS density<0.5%) and thus weak effects of BGS. This analysis reveals a substantial 5% decrease of neutral diversity near amino acid changes (Fig. 1A; randomization test $P = 6\times10^{-3}$; Methods). As expected under frequent positive selection, the decrease is more pronounced in low recombination regions (<1 cM/Mb), where the decrease of diversity is 8% on average ($P=1.5\times10^{-2}$). The decrease is stronger in the Asian (9.5%, $P=1.2\times10^{-2}$) and European (9.5%, $P=1.2\times10^{-2}$) populations than in Africa (5%, $P=7.5\times10^{-2}$) (Fig. 1B, C, D).

When we include regions of higher conserved coding density (>0.5%), as well as those located more than 1cM away from any amino acid substitution, we fail to detect any decrease in neutral diversity near amino acid substitutions. In fact, we find the opposite pattern of, on average, 4% higher diversity near amino acid substitutions ($P=4.6\times10^{-2}$), reminiscent of the results of Hernandez *et al.* (Hernandez et al. 2011). This suggests that BGS can indeed obscure signatures of positive selection in the human genome, making it essential to control for BGS and reduce its effects as much as possible when searching for positive selection.

*The functional-vs-nonfunctional test.*

The regions of low BGS in the near-vs-far test above correspond to ~30% of all the regions in which the test can be applied in principle, and ~17% of the genome in total (290 Mb of "near" and 236 Mb of "far" windows; supplemental Table 1). We are thus unable to apply this test to the majority of the genome. In addition, the choice of the threshold of CCDS < 0.5% is somewhat *ad hoc* and was driven by the need to have enough windows for the bootstrap procedure while reducing the effect of BGS as much as possible.

In order to find additional signatures of positive selection that are less sensitive to BGS and can be applied to more of the genome, we modify the near-vs-far test to compare windows that have the same overall number of amino acid substitutions (i.e. all the windows are "near"), and then contrast the windows that differ by the presence or absence of predicted *functionally consequential* substitutions, as defined by Polyphen2 (Adzhubei et al. 2010) (Methods). We reason that predicted functionally consequential substitutions are more likely to be adaptive than predicted neutral ones and should be associated with a more pronounced reduction of neutral polymorphism in their vicinity. At the same time, controlling for the total overall number of nonsynonymous substitutions in a window naturally controls for the variation in BGS.

We compare neutral diversity in 500kb windows either near predicted functional amino acid substitutions (<0.1 cM) or near predicted neutral amino acid substitutions (<0.1 cM from a neutral substitution and >0.5 cM from a functional one). The matching windows must have the same (plus/minus one) total number of amino acid substitutions. In addition, we again control for the key genomic variables (densities of coding, conserved coding and non-coding sequences, recombination rate, and GC content; Methods). In total, this functional-vs-nonfunctional test includes 823 Mb near functional substitutions and 768 Mb near nonfunctional ones (~50% of the genome in total; supplemental Table 1) and therefore greatly extends the span of the human genome we are able to analyze.

Using this test we find that neutral diversity is decreased by ~3% on average near functional compared to nonfunctional amino acid substitutions (Fig. 2A). This decrease is statistically significant but marginally so (randomization test $P=4.8 \times 10^{-2}$). As expected, the decrease of diversity is more pronounced in regions with low rates of recombination (5% on average, $P=2 \times 10^{-2}$; <1 cM/Mb) (Fig. 2A). The decrease is again weaker in the African population (3%, $P$=0.1) compared to Asian (5%, $P$=4.5x10$^{-2}$) and European populations (7%, $P$=5x10$^{-3}$) (Fig. 2B, C, D).

*A priori* we expect this test to lack power because Polyphen2 is likely to generate a substantial rate of false positives and false negatives in the identification of functional substitutions. It is extremely unlikely that all predicted functional substitutions fixed due to positive selection whereas none of the predicted neutral ones did so. The ability of this test to detect signatures of positive selection in the human genome in the face of these likely errors suggests that the rate and strength of positive selection in the human genome might have in fact been substantial.

*The omnibus near-vs-far and functional-vs-non-functional test.*
The near-vs-far test (Fig. 1) and the functional-vs-nonfunctional test (Fig. 2) search for different signals in the data and should be independent of one another. Indeed, all regions in the functional-vs-nonfunctional test are located *near* an amino acid substitution, and thus they are all in the "near" category in the near-vs-far test. The fact that the "near" regions have lower diversity than the "far" regions should not affect the results of the test that looks only within the "near" regions. In addition, we confirm by simulation that the finite number of regions used in the bootstrap procedure does not generate spurious correlations between the two tests (Supplemental Material).

The independence of the two tests allows us to combine them into a single, omnibus test and calculate a joint *P*-value (Fig. 3). In all human populations the observed combined decreases are highly statistically significant, as shown by the *P*-values of the combined randomization test in Fig. 3 (all populations combined $P$=3x10$^{-4}$; Asian, $P$=2x10$^{-4}$; African, $P$=7x10$^{-3}$; European $P$=2x10$^{-4}$). Even in Africa where the signal of positive selection is consistently weaker in both the near-vs-far and the functional-vs-nonfunctional tests, the probability of both observed decreases by chance is less than 1%. In European and Asian populations the same probability is lower than 0.1%. Taken together, these results strongly suggest that positive selection has significantly decreased neutral diversity in the human genome.

*Extreme values of XPEHH and iHS near and far from nonsynonymous substitutions.*
Both positive selection and BGS are expected to reduce the overall level of neutral polymorphism. In contrast, only positive selection, and not BGS, is expected to drive individual haplotypes to unusually high frequencies. Therefore the tests based on the presence of unusually frequent and long haplotypes, such as *iHS* (Voight et al. 2006) and *XPEHH* (Sabeti et al. 2007), should be insensitive to BGS and thus provide a less confounded approach for the systematic detection of positive selection in the genome.

We first use extensive forward simulations to confirm this intuition. We use SLiM (Messer 2013) to simulate 4 Mb regions that include a 100 kb central region where deleterious mutations occur with a predefined strength of selection and rate (Supplemental Material). We analyze a range of distributions of selective effects of deleterious mutations (Supplemental Fig. 1), including a gamma distribution that matches our best current estimate of the distribution of fitness effects of functional mutations in the human genome (Keightley and Eyre-Walker 2007). As expected, BGS has a strong effect on levels of diversity (Fig. 4) but has no detectable effect on *XPEHH* and only marginal effect on *iHS*. In the case of *iHS*, BGS slightly decreases the variance, thereby making scans for extreme values of *iHS* conservative.

We modify the near-vs-far test by using extreme values of *iHS* and *XPEHH* instead of overall levels of neutral diversity near and far from amino acid substitutions as a measure of positive selection. For *iHS*, we consider the distribution of absolute values to capture adaptation driven by both ancestral and derived alleles and to avoid issues due to potential mispolarization. Specifically, we compare the average of the top 10%, 5%, 2% and 1% *XPEHH* and *iHS* windows near and far from amino acid substitutions (Fig. 5). We use values of *iHS* and *XPEHH* calculated for the HGDP panel by Pickrell *et al*. (Pickrell et al. 2009). As before, the "near" windows are less than 0.1 cM and the "far" windows are more than 0.5 cM from any amino acid substitution. We control for levels of recombination and coding density in the bootstrap procedure. The significance of the differences between near and far windows is again calculated using the randomization test (Methods).

Fig. 5 shows clear signatures of positive selection in the *iHS* and *XPEHH* modification of the near-vs-far test. *iHS* shows significantly more extreme values near amino acid changes in all three tested populations (Fig. 5, upper row). In line with our prediction this pattern is more pronounced in low recombination regions (<0.5 cM/Mbp) (Fig. 5, right side of histograms), especially in the African population. In order to increase the statistical power, we also compare the maximum values of *iHS* in East Asians and Europeans in each window near and far from amino acid changes. Indeed, the differences of maximum *iHS* values near versus far from amino acid changes in this test are even more strongly statistically significant.

Results are essentially the same using the *XPEHH* modification of the near-vs-far test (Fig. 5, second row). We choose the ancestral, African population as the reference population and apply the *XPEHH* modification of the near-vs-far test in the two remaining populations. The results are significant in both populations and again become more pronounced in the low recombination regions (<0.5 cM/Mbp) and when the two populations are combined.

Because *iHS* and *XPEHH* are insensitive to BGS we were able to carry out these tests even in regions that have high coding density and in which the tests that rely on the overall level of polymorphism would therefore be too biased by BGS against the detection of positive selection. Specifically, the "near" windows in the *iHS* and *XPEHH* tests represent a total of 1.56 Gb and "far" windows represent a total of 618 Mb, extending the amount of sequence used for the detection of signatures of positive selection to ~70% of the human genome (Supplemental Table 1).

*The CLR test of positive selection*

The previous tests suggest that positive selection is more common near compared to far from amino acid substitutions. As a consequence, the allele frequency spectra of neutral polymorphism should more often show characteristic deviations consistent with positive selection near amino acid substitutions. We test this prediction using the composite likelihood ratio test (*CLR*) (Williamson et al. 2007). The *P*-values of the *CLR* test were retrieved from Williamson *et al*. (Williamson et al. 2007) for the East Asian and European populations. We do not consider the results of the *CLR* test for the African population because they were calculated by Williamson *et al*. (Williamson et al. 2007) using a sample of strongly admixed African-Americans individuals (Note that running the *CLR* test on the 1,000 genomes phase 1 data would have been computationally prohibitive). We take the lowest *P*-value found in each window and then compare the average 10%, 5%, 2% and 1% lowest *P*-value windows near and far from amino acid substitutions as above. We do detect more extreme values of *CLR P*-values near amino acid substitutions in Europeans, in the low recombination regions (<0.5 cM/Mbp) in East Asians, and in all regions in the combined analysis of the European and East Asian populations (Fig. 5 lower row). These results again confirm that adaptation does appear to be more common near compared to far from amino acid substitutions in the human genome.

*Forward simulations of positive selection.*

We use SLiM (Messer 2013) to run forward simulations of positive selection (Methods) in order to determine how strong and frequent recurrent selective sweeps need to be in order to decrease neutral diversity near amino acid substitutions between 2% and 9.5% in 500 kb windows, as observed in the data (Figs. 1 and 2). In particular, we focus on regions of low recombination (<1 cM/Mb) and simulate adaptation with three different rates of adaptive amino acid substitutions (proportion of substitutions that are adaptive $\alpha$ =10, 20 and 40%) and two different selection regimes (selection coefficient $s$ = 0.01 and 0.05). Surprisingly, the amount of strong positive selection needed to explain the observed reduction in diversity is very high (Fig. 6). The observed 9.5% reduction in Europe and Asia is similar to the average reduction expected if 40% of amino acid substitutions were adaptive with a selection coefficient of $s$=0.05; they are in the high range if 10 or 20% of amino acid changes were adaptive with $s$=0.01.

This rate appears higher than that estimated with MK approaches, which predict that approximately 20% or fewer of amino acid changes (Boyko et al. 2008; Messer and Petrov 2013) were adaptive in the human lineage. The MK estimate includes both strongly ($s$>0.01) and weakly ($s$<0.001) selected substitutions, whereas we infer that at least 10% of the amino acid substitutions were driven by strong selection ($s$>0.01). This implies that either at least half or more of the adaptive amino acid substitutions were driven by strong selection, or, alternatively, that the majority of adaptive changes are not amino acid substitutions themselves, but instead are adaptations at nearby, possibly regulatory, sites.

*Adaptation is centered at the ENCODE-defined regulatory elements*

The above simulations suggest that adaptation by amino acid substitutions is unlikely to generate all of the observed signatures of adaptation. We therefore search for adaptation at regulatory regions by focusing on the ENCODE-defined regulatory elements (ERE) (Gerstein et al. 2012). We examine the correlation between the density of ERE and *iHS* in three populations of the 1,000 Genomes phase 1 project (Abecasis et al. 2012) (Supplemental Material). ERE density in our analysis is the density of elements predicted as DNASEI hypersensitive sites and also as transcription factor binding sites identified via Chip-Seq by the ENCODE Consortium (Gerstein et al. 2012). In Europe and Asia, absolute values of *iHS* correlate positively with ERE density (Fig. 7B,C), but the correlation is more subtle in Africa, where it becomes positive only in low recombination regions (Fig. 7A). The correlation is notably stronger in regions with low recombination rates, as expected under frequent positive selection.

However, ERE density also correlates strongly with coding density ($n$=43,780, Spearman's $\rho$=0.73, $P$<2x10$^{-16}$), and coding density correlates with *iHS* (Fig. 7). In order to disentangle the respective contributions of coding and regulatory sequences on the observed signal of recent positive selection, we calculate the reciprocal partial correlations between (i) *iHS* and ERE density controlling for coding density and (ii) *iHS* and coding density controlling for ERE density. When using the whole genome regardless of recombination, the partial correlations between *iHS* and ERE or coding density are weak and inconsistent between different human populations, being either positive in Asia or negative in Africa (Fig. 7A,B,C). In low recombination regions (<0.5 cM/Mb), where the effects are expected to be the strongest and clearest, the results are striking: while the partial correlation between *iHS* and ERE density appears virtually independent from coding density, the correlation between *iHS* and coding density disappears entirely once controlling for ERE density. This result provides strong evidence that most signals of positive selection in the human genome are indeed due to adaptation centered in regulatory rather than in coding sequences.

**Discussion.**

In this study, we have used a number of independent approaches to detect and quantify the effects of positive selection on patterns of variation in the human genome. Our results show that positive selection was frequent in the human lineage, but that its effects are challenging to detect given that BGS masks some key signatures of recurrent positive selection. Specifically, the key prediction of recurrent and pervasive positive selection is that neutral polymorphism should be lower in regions with more functional, for instance, nonsynonymous substitutions while controlling for the overall functional density (Cai et al. 2009). Perhaps counterintuitively, BGS is expected to generate precisely the opposite signature: regions of the genome that have high functional density but very few nonsynonymous substitutions are likely to be under stronger constraint and thus should exhibit stronger BGS and lower levels of neutral polymorphism. This means that the standard approaches that search for adaptation using the signature of low levels of polymorphism next to nonsynonymous substitutions, such as those of Macpherson *et al.*, Cai *et al.*, and Sattath *et al*. (Andolfatto 2007; Macpherson et al. 2007; Cai et al. 2009; Sattath et al. 2011), are likely to underestimate the effect of positive selection.

This underestimation is likely to be marginal in small and functionally dense genomes, such as that of Drosophila, where levels of BGS are expected to be homogeneous along the genome. However, in larger genomes with heterogeneous distribution of functional sequences, such as that of humans, the levels of BGS vary sharply along the genome and this bias against finding signatures of positive selection can become profound.

We first tested whether stronger BGS far from functional substitutions indeed masks signatures of positive selection in the human genome. Specifically, we compared levels of neutral polymorphism near and far from amino acid substitutions in regions with matching functional densities, GC contents, and recombination rates. We conducted this test separately in the regions with low functional density, and thus overall low levels of BGS, and in the entire genome. This test revealed lower levels of polymorphism near amino acid substitutions in regions of low functional densities, while showing higher levels of polymorphism near amino acid substitutions in the genome as a whole. This suggests that the bias towards stronger BGS far from amino acid substitutions does hide signatures of positive selection genomewide. It also predicts that positive selection should be detectable if we use signatures of positive selection insensitive to BGS.

Although BGS has strong effects on the overall levels of polymorphism, it is unlikely to mimic other signatures of positive selection, such as the presence of long and frequent haplotypes driven into the population by selective sweeps. We conducted extensive simulations of BGS under varying rates and patterns of deleterious mutation and showed that tests of selection based on the presence of such long and frequent haplotypes (*iHS* and *XPEHH*) are indeed virtually insensitive to BGS. The only detectable effect is that *iHS* becomes marginally conservative in that it is somewhat less likely to exhibit extreme values under neutrality and BGS.

As expected under pervasive positive selection, we detected significantly more extreme values of *iHS* and *XPEHH* near amino acid substitutions. Because these statistics are insensitive to BGS we were able carry out this analysis systematically on a genomewide scale, without having to restrict it only to regions with low functional density. Moreover, we confirmed that the regions near amino acid substitutions have skewed allele frequency spectra consistent with positive selection.

All the evidence together argues strongly that positive selection left detectable effects on patterns of variation in the human genome. However, it is also clear that these patterns are difficult to detect, both because BGS systematically hides these signals and also because a number of other processes affect levels of polymorphism across the human genome. In this study, we carefully controlled for this variation by always comparing windows that were matched by all presently known factors associated with levels of polymorphism: functional density and levels of constraint (and thus BGS), recombination rate, GC content, mutation rate (by controlling for the levels of divergence at neutral sites with macaque), sequencing read depth, and by only measuring levels of polymorphism in nongenic, unconstrained regions that are free from repeats. It is of course possible that there are yet unknown genomic variables that correlate with levels of neutral polymorphism, but we believe that it is extremely unlikely that (i) they are uncorrelated with other genomic variables that we have controlled in this study and (ii) that they would correlate both with levels of polymorphism and generate unusually long and frequent haplotypes detected by *iHS* and *XPEHH*.

Demographic perturbations such as bottlenecks and admixture can generate additional variability in levels of polymorphism and haplotype structure. However, it is hard to imagine a scenario in which these demographic perturbations would affect windows near amino acid substitutions differently from those that are far from amino acid substitutions in the long history of evolution since divergence of humans and chimpanzees. First, the vast majority of the amino acid substitutions happened long ago, prior to any demographic event in question. Second, the windows near and far from amino acid substitutions that are used in the comparisons have had exactly the same demographic history. Thus the main effect of demography is to increase variance in levels of polymorphism both in windows near and far from amino acid substitutions, but it is unlikely to generate false positives by itself.

It is worth noting that although signals of positive selection are detectable in all tested populations, these signals are systematically stronger in the out-of-Africa populations. On possible explanation for this is that the masking effects of BGS are stronger in Africa then out-of-Africa**.** This should reduce the signal of positive selection in the near-versus-far test that uses levels of neutral polymorphism. It is also possible that there have been more recent sweeps in out-of-Africa populations, possibly due to the need for adaptation to new environmental challenges as proposed previously (Williamson et al. 2007).

We next sought to quantify how much strong positive selection is needed to explain our results. We showed that if adaptation only happened at nonsynonymous sites, then a scenario in which 10% of amino acid substitutions are adaptive with $s$=0.01 is the lower limit of the rate and strength of adaptation in recent human evolution. Given that McDonald-Kreitman tests estimated that at most ~10-20% of amino acid substitutions are advantageous (Boyko et al. 2008; Messer and Petrov 2013), this result would make sense either if all amino acid substitutions were strongly advantageous, or if many adaptations took place at nearby regulatory sites.

If ~10% of all adaptive substitutions are strongly advantageous in humans, comparable to what was estimated in Drosophila (Macpherson et al. 2007; Sattath et al. 2011), then ~10 times as many adaptations must have taken place at regulatory sites. If the proportion of adaptation driven by strong positive selection is lower, then the proportion of regulatory changes responsible for adaptation increases even further. We provided evidence for the assertion that much adaptation is driven by regulatory changes by demonstrating that signatures of recent and strong adaptation correlate much better with the density of ENCODE regulatory elements (Gerstein et al. 2012) than with the density of coding sequences, despite the fact that the latter is much less noisy.

Our lower estimate for the rate of adaptation in the human genome is one adaptive substitution per 1,000 years if all adaptations were driven by strong selection ($s = 0.05$) and if we assume that humans diverged from chimpanzees 5 MYA and had a generation time of 25 years since then. Over the past one hundred thousand years, we therefore expect ~100 strong adaptive substitutions. Given that this is roughly the time over which scans for selection have power to detect true positives (Przeworski 2002; Sabeti et al. 2006), our estimates suggest that genome scans represent a valuable avenue for the study of human adaptation (Grossman et al. 2010). However, given that cumulatively scans for selection detected thousands of candidate adaptive loci (Akey 2009), the scans either suffer from a substantial rate of false positives, or many adaptations they detect were driven by weaker ($s<0.05$) selection or were either partial or soft sweeps.

Our results establish that advantageous mutations have been frequent during recent human evolution and that many adaptive changes may have been strongly beneficial. We argue that the majority of adaptive changes are located in regulatory sequences, providing confirmation of the King and Wilson hypothesis (King and Wilson 1975) that most adaptive divergence is regulatory and not coding. The challenge for the future is to identify these human-specific adaptations and to understand the role they played in human evolution.

## Methods

**Human-specific nonsynonymous and synonymous fixed substitutions**

Human-specific nonsynonymous and synonymous substitutions were obtained using human-chimpanzee-orangutan coding DNA sequence (CDS) alignments. Human CDS are first extracted from the Ensembl v64 database (http://www.ensembl.org/). For each gene, only the longest CDS is retained. Human longest CDS are then mapped onto the chimpanzee and orangutan genomes using Blat (Kent 2002) (protein-protein Blat, 60% minimum identity). The best, highest identity chimpanzee and orangutan Blat hit sequences are then mapped back on the human genome. Only those human-chimpanzee and human-orangutan best reciprocal hits are retained for further analysis. Extracting chimpanzee and orangutan CDS from their respective genomes using Blat instead of directly using Ensembl annotations ensures that the sequences used during subsequent global alignment steps have good local similarity. The analysis is further restricted to those best Blat reciprocal hits that coincide with Ensembl v64 one-to-one orthologs. A total of 17,237 CDS multiple alignments are finally obtained using Prank (Loytynoja and Goldman 2008) under the codon evolution model settings. Prank used with its codon evolution model was previously shown to be the most accurate solution to align CDS (Fletcher and Yang 2010). From these alignments, a total of 27,538 and 40,709 nonsynonymous and synonymous human-specific substitutions are identified, respectively. This includes only those cases where chimpanzee and orangutan both exhibit the same nucleotide at the orthologous position. Of the 27,538 nonsynonymous substitutions, a total of 21,278 are fixed in all African, Asian and European populations. Of the 40,709 synonymous substitutions, 32,666 are fixed. The ratio of the number of fixed nonsynonymous to fixed synonymous substitutions is 65.1 %, which is in very good agreement with the previous result of 64% obtained by Boyko *et al.* (Boyko et al. 2008). Only diversity patterns close to fixed substitutions are analyzed in the near versus far and the functional versus non-functional tests. Focusing on fixed substitutions is therefore intended to make results easier to interpret. This is also expected to be conservative when searching for sweeps, because we exclude fixations that occurred after the split of African and non-African populations.

**Polyphen2 analysis**

We use Polyphen2 (Adzhubei et al. 2010) to identify which human-specific amino acid substitutions are more likely to be functionally consequential. Polyphen2 annotates SNPs but can also be used to annotate fixed amino acid changes by using the REVERSE option. Of the 21,278 fixed amino acid changes specific to the human lineage, 18,924 (89%) can be annotated. Of these, 15,488 are annotated as benign, 1,874 as possibly damaging, and 1,562 as probably damaging by Polyphen2. The possibly damaging and probably damaging amino acid changes (18% of the total) are more likely to be functionally consequential than the benign ones. Thus, in the functional versus non-functional test (Main text and Bootstrap procedure below), functional windows are those close to a possibly or probably damaging amino acid change and the non-functional windows are those close to a benign amino acid change, but far from any possibly or probably damaging one.

**Neutral diversity**

Neutral diversity is measured using average heterozygosity $\pi$, measured as $2f(1-f)n/(n-1)$ where $f$ is the frequency of the non-reference allele in the 1,000 Genomes phase 1 20100804 release (December 2010 update) and $n$ is the number of chromosomes in 500 kb windows (see below for an in-depth discussion on window size). More specifically, average heterozygosity is calculated separately for the three African, Asian and European populations. We use only positions outside of CDS, UTRs (from Ensembl v64) and phastCons CNEs (from the UCSC Genome Browser), simple repeats and transposable elements identified by Repeatmasker (http://genome.ucsc.edu/). Excluding functional elements, repeats and positions not aligned with a nucleotide in macaque, approximately a third of the positions within windows can be used on average to measure neutral diversity. We also exclude all windows closer than 5 Mb to centromeres or telomeres from our analysis. Diversity is further scaled by the number of positions found to be divergent between human and macaque in human-macaque Blastz (Schwartz et al. 2003) alignments retrieved from the UCSC Genome Browser (http://genome.ucsc.edu/). This is done to eliminate the effect of local variations in mutation rate or remaining strong selective constraint. Because local changes in mutation rate and strong selective constraint affect both diversity and divergence equally, using the ratio of diversity on divergence removes at least partially the effects of heterogeneous mutation rates and selective constraint. Using scaled diversity implies that only those positions where a nucleotide (non-N or any other undefined position) is aligned with a nucleotide in macaque are used.

**Defining genomic windows to measure diversity**

Scaled neutral diversity is calculated within 500 kb windows sliding every 5 kb in the genome. A fixed physical size is chosen instead of a genetic size in order to make the windows used in the near versus far and the functional versus non-functional tests comparable. In fact an important problem with using windows with a fixed genetic size is that they can vary greatly in physical size. Depending on the recombination rate, a 0.1 cM window in the human genome can represent a physical size of 50 kb (if the recombination rate is 2 cM/Mb) or a megabase (if the recombination rate is 0.1 cM/Mb). Using a fixed genetic size can thus result in using windows with vastly different absolute content of functional elements. For instance background selection (BGS) is expected to decrease diversity more strongly in a 0.1 cM, one megabase window with 1% (10,000) of its positions in coding exons compared to a 0.1 cM, 50kb window also with 1% (500) coding exon positions. We therefore choose to use fixed physical distances and to match recombination rates as part of the bootstrap procedure used for our near versus far and functional versus non-functional tests (see Bootstrap procedure below). We use large windows of 500kb to prevent other additional issues with using smaller windows. First, bigger windows tend to exhibit less variable, closer to genomic average parameters such as GC content, CDS, UTR content, and others compared to smaller windows. This is crucial for the bootstrap procedure used in the near versus far test and in the functional versus non-functional test. Because large windows tend to be closer to the genomic average compared to smaller windows, in both tests it is much easier to find control windows that match the tested window in terms of recombination, GC content and diverse functional contents (see Bootstrap procedure below). For instance, using 500 kb windows for the near versus far test (conserved coding density<0.5%, windows further than 1 cM windows excluded, recombination rate lower than 1 cM/Mb), after 10 bootstraps we can match on average 3,260 near windows with far windows out of the 13,678 near windows in the genome. In other words, 24% of the windows of interest can be controlled for. Using 100kb instead of 500kb windows, we found that only 1.8% of the near windows can be used.

       Second, an important issue with small windows is that functional elements outside of the windows but at their immediate proximity may influence diversity inside

windows more strongly than if larger windows are used. Consider for example a 100 kb window within a region of low recombination rate. This window has a 1% CDS density, but is surrounded by regions with a 5% CDS density. In such a case it is likely that BGS due to the surrounding CDS affects neutral diversity within the window even more than the CDS within the window itself. Using larger windows does not remove this edge effect, but it does improve it to some extent by reducing the effect of outside compared to inside functional elements on diversity. The effect of nearby functional elements on diversity can be estimated by measuring the partial correlation between neutral diversity and the amount of functional elements surrounding windows at a close genetic distance, controlling for the amount of functional elements within windows and recombination rate. For CDS, we measure that using 100kb windows sliding every 50kb, the partial Spearman's correlation coefficient is -0.14 between diversity in the African population and the absolute amount of CDS surrounding the windows up to a genetic distance of 0.1 cM ($n$=44,986, $P<2\times10^{-16}$). The same partial correlation coefficient is reduced to -0.06 when using 500 kb windows sliding every 50kb ($n$=44,858, $P<2\times10^{-16}$). The smaller influence of nearby functional elements on bigger windows reflects the fact that on average any position in bigger windows is further from the surrounding functional elements compared to smaller windows. Within 500kb windows, the average physical distance of a position to the closest window boundary is 125 kb, whereas for 100 kb the average distance is only 25 kb.

Finally, using larger windows makes the measures of diversity less noisy, especially given the fact that on average only a third of the positions within each window are used to measure scaled neutral diversity (as a reminder, those positions that occur out of CDS, UTR and CNEs, out of repeats and that are aligned with a nucleotide in macaque).

**Bootstrap procedure**

In humans local functional density is very heterogeneous and is a main determinant of neutral diversity. Regions of high functional density have higher levels of BGS and hence lower levels of neutral diversity (McVicker et al. 2009; Lohmueller et al. 2011). GC content and recombination also have a strong influence on levels of neutral diversity (Results). In our study we want to characterize the effect of positive selection on neutral diversity. This is done by comparing neutral diversity in regions of the genome where the rate of positive selection is expected to be higher with neutral diversity in regions where the rate of positive selection is expected to be lower. Genomic windows with potentially higher rates of positive selection are called tested windows, and genomic windows with potentially lower rates of positive selection are called control windows. In the near versus far test, tested windows are the windows near amino-acid changes (nearest amino acid change at less than 0.1 cM from the center of the window) and the control windows are windows far from any amino-acid change (>0.5 cM). In the functional versus non-functional test, tested windows are the windows near functional amino acid changes according to Polyphen2 (<0.1 cM) and the control windows are windows near non-functional amino-acid changes (<0.1 cM) but far from any functional amino-acid change (>0.5 cM). In addition to positive selection, we also tested whether windows very far from any amino acid change (>1 cM) experience more BGS than windows moderately far from amino acid changes (between 0.1 cM and 1 cM). In this case tested windows are the windows between 0.1 cM and 1 cM and control windows are the windows further than 1 cM from any amino acid change.

The major challenge when testing positive selection by comparing tested and control windows is to make sure that both kinds of windows are as similar as possible. One may think of an example where in tested windows the percentage of positions within CDS is 2% on average and only 0.5% in control windows. In this case there are four times more CDS in the tested windows than in the control windows. BGS is thus stronger in the tested windows. In such an example, neutral diversity is lower in tested windows than in control windows not because of positive selection but because of stronger BGS, and it is impossible to conclude anything about positive selection. This example shows that in order to be conclusive about positive selection we need to compare windows with levels of BGS as similar as possible. This means that the tested and control windows need to have on average similar functional densities, in addition to similar recombination rates and GC content. This is achieved by using a simple bootstrap procedure. For each tested window, we match a control window whose characteristics are not more different than fixed thresholds compared to the tested window. These characteristics are the average recombination rate in the window obtained from the most recent decode 2010 genetic map (Kong et al. 2010), GC content, CDS density (Ensembl v64), conserved coding sequences (CCDS) density (Ensembl v64), UTR density (Ensembl v64), and total functional density (TFD). CCDS are the 83% of coding sequences that overlap conserved segments (mammal-wide and/or primate-wide) predicted by phastCons (Siepel et al. 2005) and available at the UCSC Genome Browser (phastCons applied to a genome alignment of 44 mammals). TFD is the percentage of positions in a window that are in at least one of these different types of functional elements: CDS, CCDS, UTR, phastCons conserved non-coding element (CNE). In addition, we also control for the amount of surrounding CDS, which is the number of positions within a CDS up to 0.1 cM upstream and 0.1 cM downstream of a window.

For each tested window, we find a matching control window whose recombination, GC content, CDS, CCDS, UTR, TFD and surrounding CDS are comprised between x% and y% of their values in the tested window. The values of x and y are specific to each of the controlled factors, and x is smaller than one while y is greater than one. For example we could ask control windows to have a CDS density comprised between x=80% and y=120% of the tested window CDS density. In practice we adjust the thresholds so that when the bootstrap is complete tested windows and control windows have a very similar average recombination rate, average GC content, average CDS, CCDS, UTR, TFD and surrounding CDS. In addition, we also make sure that they have very similar phastCons CNE density.

Although we cannot avoid slight differences, we make sure they are in the conservative direction. For example the average CDS density in the control windows may be 3% higher than in the tested windows, and the average recombination rate may be 5% lower. When no matching control window is found in the genome the tested window is excluded from the analysis. The same control window can be used several times as a match for several tested windows. The different amounts of sequences that could be used for each test are shown in Supplemental Table 1. The x% and y% thresholds used for the different tests conducted in this analysis are provided in Supplemental Table 2. Note that the thresholds were adjusted so that they could be used for all the repetitions of a given test in various conditions. For example in the near versus far test we used thresholds that are adapted whether or not we use only windows below a fixed recombination threshold, and whether or not we use only low CCDS windows (Results). This is to ensure that the results obtained under these different conditions can be fairly compared between each other.

For each test the bootstrap procedure is conducted ten independent times. Each time we calculate the average neutral diversity in tested windows $\Pi_{tested}$, the average neutral diversity in control windows $\Pi_{control}$ and the ratio $\Pi_{tested}/\Pi_{control}$. Different realizations of the bootstrap procedure give very similar $\Pi_{tested}/\Pi_{control}$ ratios. For all tests and for each realization the ratio $\Pi_{tested}/\Pi_{control}$ never differs by more than 10% of its average over the ten realizations. The observed ratios $\Pi_{tested}/\Pi_{control}$ shown in Figs. 1 and 2 represent the average over the ten realizations of the bootstrap procedure. Because there is so little variation between the different realizations of the bootstrap procedure, we always use the first realization for running populations simulations (see Population simulations below) and for calculating *P*-values of the randomization test (see Randomization test below). Note also that we do not include average sequencing depth in the windows as one of the controlled variables although it is well known to have an effect on the estimation of neutral diversity. This is because we found this is not necessary since on average the tested and control windows retained by the bootstrap procedure have extremely similar average sequencing depths that never vary by more than 0.5% from each other.

**Randomization test**

We use a randomization test to estimate the significance of the differences of neutral diversity we observe between tested and control windows used in the bootstrap procedure. In order to obtain a random distribution of $\Pi_{tested}/\Pi_{control}$ for a given realization of the bootstrap procedure, we need to shuffle tested and control windows while accounting for a number of features of the analysis. First, the tested and the control windows are often clustered together, much like the windows represented along a chromosome in Supplemental Fig. 3. $\Pi_{tested}$ and $\Pi_{control}$ are calculated from groups of neighboring, overlapping windows that have correlated neutral diversity values. Compared to a situation where we would have the same number of windows but all independent from each other, this grouping substantially increases the variance of $\Pi_{tested}$, $\Pi_{control}$ and thus of the ratio $\Pi_{tested}/\Pi_{control}$. Shuffling individual windows independently from each other is therefore very likely to greatly underestimate the true variance of the ratio. Second, during the bootstrap procedure the same control window can be matched with several tested windows, which should also be taken into account during the randomization process. In order to maintain the structure of the sampling scheme used in the bootstrap procedure, we shuffle blocks of neighboring windows (Supplemental Fig. 2). Windows used in the bootstrap procedure are first ordered according to their genomic positions. We then cut 20 segments of equal size (Supplemental Fig. 2 represents a situation with only three segments). This is done to maintain the grouping of windows. The 20 segments are then shuffled to obtain a new random ordering of windows. In addition a segment can be flipped with a probability of 50%. The same sampling scheme that was used during the bootstrap procedure is finally applied to the randomized windows. For example in the genome the positions 19,20 and 21 are occupied by tested windows tested_19, tested_20 and _tested_21 that are all matched with the same control window control_29 at position 29 (Supplemental Fig. 2). After the randomization, positions 19, 20 and 21 are now occupied by tested windows tested_8, tested_9 and tested_10 that now all match with window tested_18 at position 29. This way the neighboring windows tested_19, tested_20 and tested_21 have been replaced by three other neighboring windows, and window tested_18 matches three times as window control_29. The randomization process is repeated 10,000 times to obtain the *P*-value for the test. *P*-values are calculated as the proportion of randomizations where random $\Pi_{tested}/\Pi_{control}$ is lower or higher than the observed $\Pi_{tested}/\Pi_{control}$ depending on the case studied. This means that the randomization test is a one-sided test.

**Population simulations**

In our study we use forward simulations to estimate the ranges of the ratios of $\Pi_{near}/\Pi_{far}$ and $\Pi_{func}/\Pi_{non\text{-}func}$ under both a demographic scenario of panmixia with no advantageous mutation and under a scenario of panmixia with different rates and strengths of positive selection. Simulations were conducted using SLiM (Messer 2013). We simulate segments of the human genome where windows were sampled by the bootstrap procedure. Supplemental Fig. 3 shows how those segments are defined based on where the sampled windows are in the genome and how far they are from each other. In Supplemental Fig. 3, 500 kb sampled windows define three non-overlapping groups along a chromosome. The first and second groups (starting from the left) are at distance of 0.23 cM from each other. These two groups are fused together to form a genomic segment that includes them both. The segment is further extended 0.1 cM upstream and 0.1 cM dowstream to avoid edge effects and to include the effect of eventual neighboring advantageous mutations not included in, but close to, the sampled windows (Supplemental Fig. 3). The third group is at 0.84 cM and is treated as an independent segment. Overall, groups of windows closer than 0.5 cM from each other are fused together while groups further than 0.5 cM from each other are treated as independent simulated segments.

All the segments in the genome are simulated independently and the simulated ratios $\Pi_{tested}/\Pi_{control}$ are calculated exactly as they are using the bootstrapping procedure. This means that the same 500 kb windows are used and that within each window, variants whose coordinates fall within a functional element, a repeated element, or do not align with macaque in the real genome, are excluded from the calculation of simulated diversity. The whole operation is repeated 100 times for the estimation of confidence intervals of $\Pi_{tested}/\Pi_{control}$.

The recombination maps used in each segment match the Decode 2010 recombination map (Kong et al. 2010). The simulations were conducted using a population of 500 individuals and the recombination and mutation rates were rescaled accordingly to match the average recombination rate (1.16 cM/Mb) and the average heterozygosity (0.001) observed in the human genome. After a burn-in of 5,000 generations, the neutral simulations are continued for 1,000 additional generations (this is equivalent to 20,000 generations in a non-rescaled 10,000 individuals human population). Simulations with positive selection are continued for 2,500 generations after the burn-in to ensure that all advantageous mutations introduced after the burn-in are given a fair amount of time to fix.

For the simulations with positive selection, we introduce advantageous mutations at random generation times with a fixed rescaled selection coefficient at positions where amino acid changes are found in the human genome. As an example, we can simulate a scenario where 10% of the amino acid changes were adaptive with $s$=1%. The selection coefficient of 1% in a 10,000 individuals population is rescaled to 20% in our 500 individuals simulated population to maintain the same intensity of selection. In order to obtain 10% of fixed adaptive mutations, given the probability of fixation (2$s$=40%) we need 25% of the introduced mutations with $s$=20%. These advantageous mutations are introduced randomly among all the locations with an amino acid change. For the sake of speed in our simulations with positive selection we use 2,500 generations after burn-in although in our rescaled population the number of generations to the human-chimpanzee most recent ancestor is 10,000 generations (rescaled from 200,000 generations assuming a TMRCA of 5 My and a generation time of 25 years). Advantageous mutations were thus attributed an introduction time between 1 and 10,000 generations after burn-in, but only those mutations having a random introduction generation between 1 and 2500 were actually introduced in the population.

**Acknowledgements:**
We thank Hugues Roest Crollius (ENS Paris) for sharing his computational resources, Pardis Sabeti, Kirk Lohmueller, Hunter Fraser, Noah Rosenberg and members of the Petrov lab, especially Pleuni Pennings, Fabian Staubach, Diamantis Sellis, Rajiv McCoy, Anna-Sophie Fiston-Lavier and Nandita Garud for helpful comments on the manuscript.

**Figures legends:**

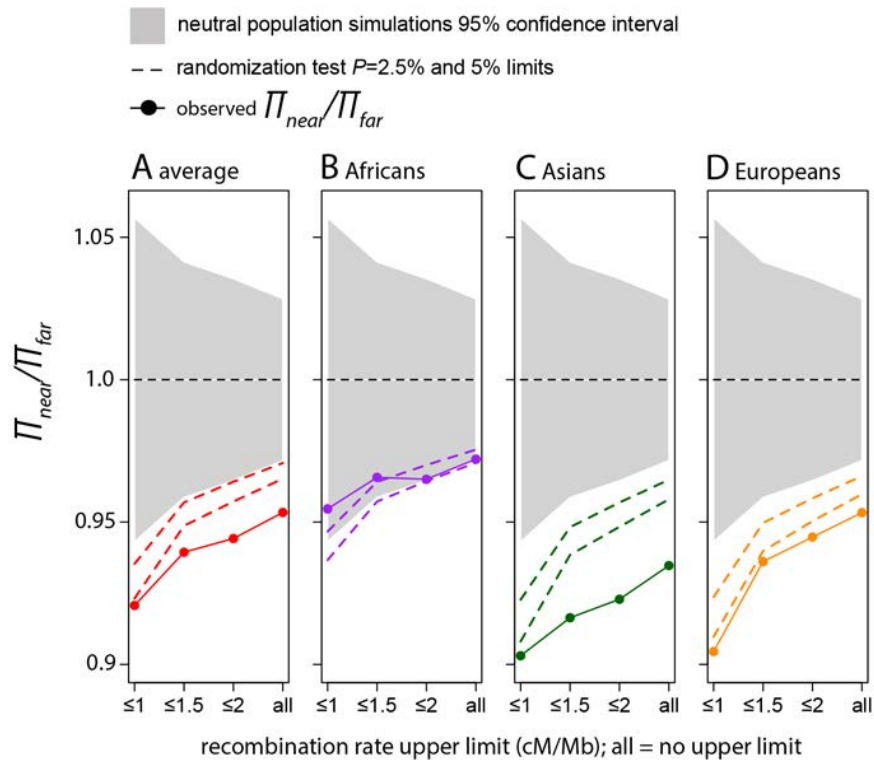

**Figure 1. Lower diversity near versus far from amino acid substitutions**
Each panel shows the level of synonynmous heterozygosity near amino acid changes ($\pi_{near}$) compared with that far from amino acid changes ($\pi_{far}$) for the particular subpopulation. $\pi_{near}$ is the average over all near windows (<0.1 cM) from the bootstrap procedure (Methods). $\pi_{far}$ is the average over all far windows (>0.5 cM but <1 cM). The grey area depicts the 95% confidence intervals based on neutral simulations (Methods). The dashed lines show 95% and 97.5% confidence intervals based on a randomization test (Methods). Randomization tests result in at most 35% greater variance than the neutral simulations. This is expected given that neutral simulations do not account for complex demography and other sources of noise in the data. On the x axis, ≤1, ≤1.5 and so on means that we use only windows with recombination rates lower or equal to 1 cM/Mb, 1.5 cM/Mb and so on to compare diversity near and far from amino acid substitutions. "all" means that we use all windows independently of their recombination rates.

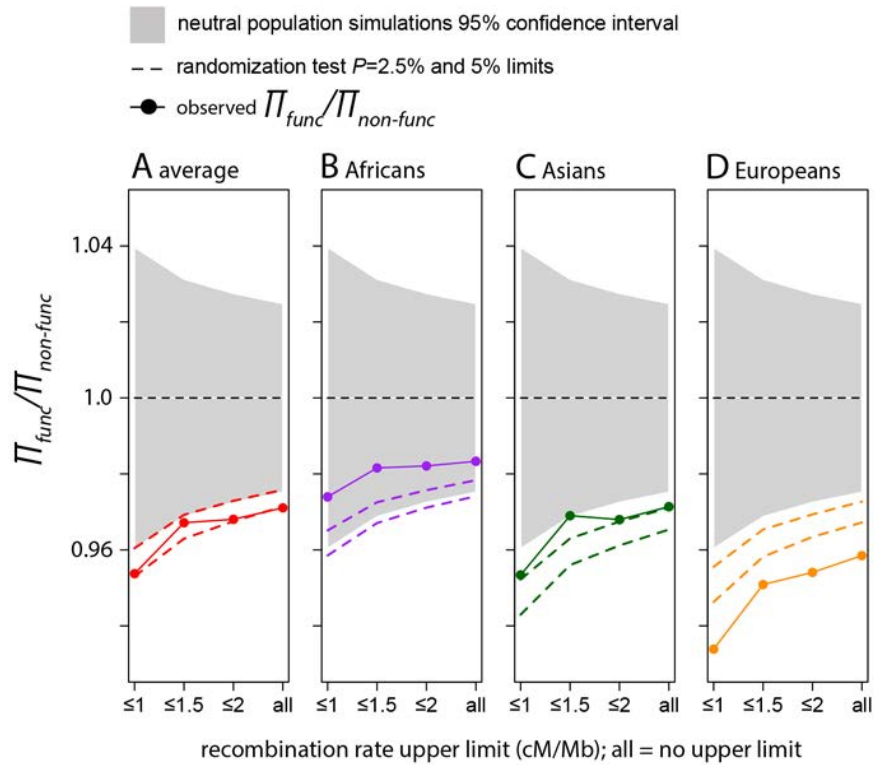

**Figure 2. Lower diversity near functional amino acid substitutions**
We compared heterozygosity near functional amino acid changes $\pi_{func}$ with heterozygosity near non-functional amino acid changes $\pi_{non-func}$. $\pi_{func}$ is the average over all functional windows from the bootstrap procedure (Methods). $\pi_{non-func}$ is the average over all non-functional windows. The grey area depicts the 95% confidence intervals based on the neutral simulations (Methods). The dashed lines show 95% and 97.5% confidence intervals established based on a randomization test (Methods). On the x axis, ≤1, ≤1.5 and so on means that we use only windows with recombination rates lower or equal to 1 cM/Mb, 1.5 cM/Mb and so on to compare diversity near and far from functional amino acid substitutions. "all" means that we use all windows independently of their recombination rates.

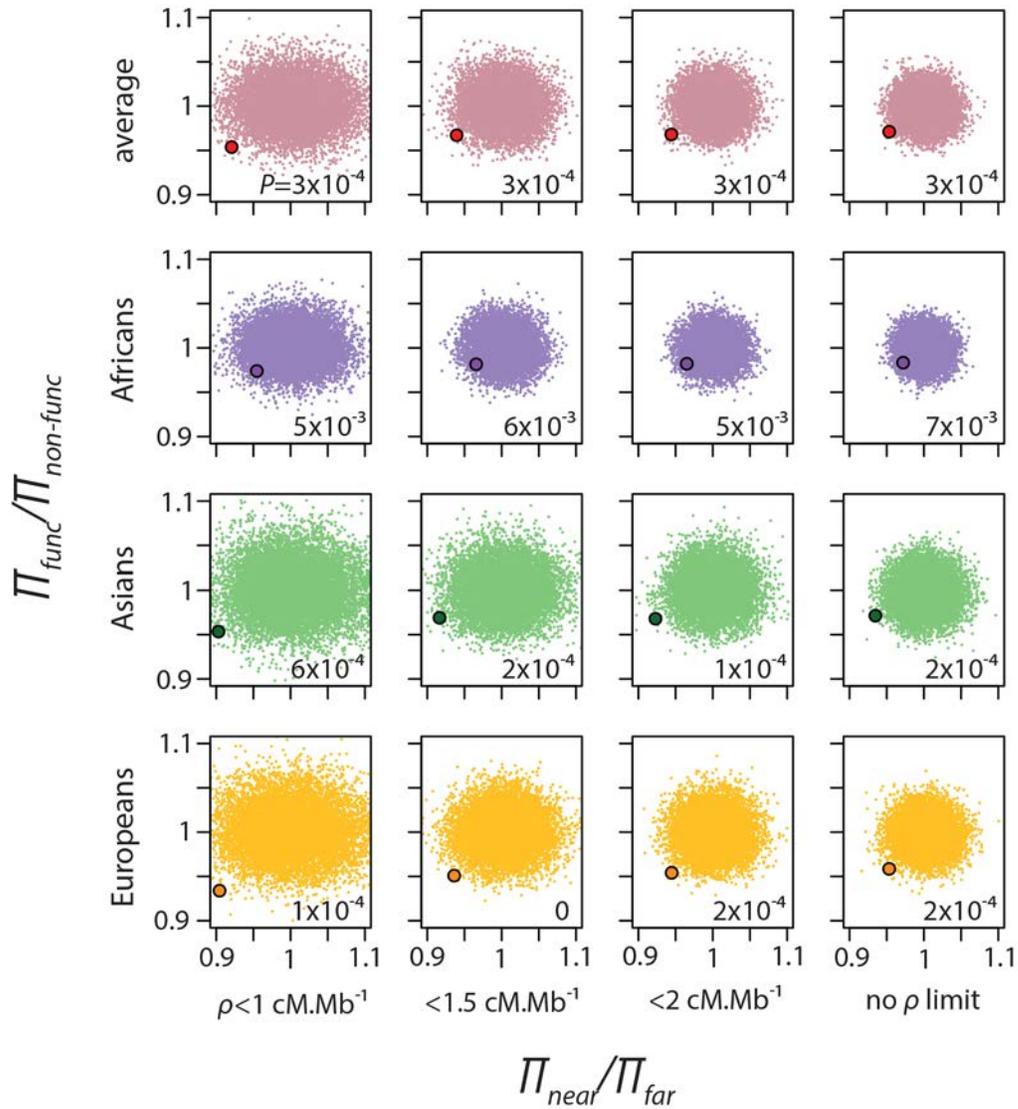

**Figure 3. Combined near versus far and functional versus non-functional tests**
Clouds of small dots represent the ratios $\pi_{near}/\pi_{far}$ and $\pi_{func}/\pi_{non-func}$ obtained with the randomization test. The larger dot in each graph represents the observed $\pi_{near}/\pi_{far}$ and $\pi_{func}/\pi_{non-func}$. The numerical values at the lower right side of each graph are the P-values obtained after 10,000 iterations of the randomization test. The P-values are estimated as the proportion of the randomizations that give values below the observed value in both tests.

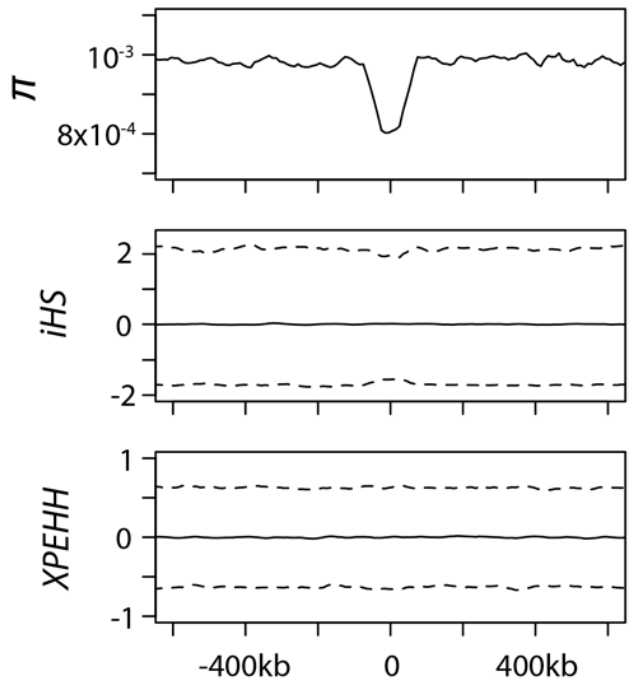

**Figure 4. Robustness of *iHS* and *XPEHH* to BGS**

We tested the effect of BGS on *iHS* and *XPEHH* (Results and Supplemental Material). Upper row: average heterozygosity. Middle row: *iHS*. Lower row: *XPEHH*. The full lines represent average *iHS* or *XPEHH* along the simulated region. The dashed lines represent the limits of *iHS* or *XPEHH* 95% confidence intervals.

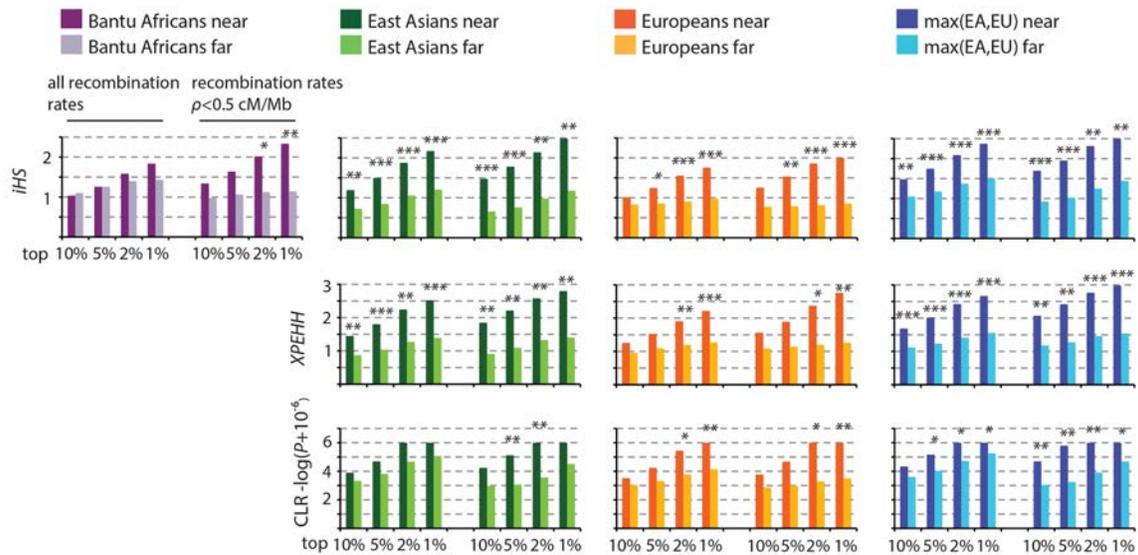

**Figure 5. Stronger haplotype and SFS based signals of positive selection near amino acid changes.**
This figure shows the comparison of the top 10%, 5%, 2% and 1% *XPEHH* and *iHS* windows near and far from amino acid changes. For the *CLR* test we take the lowest *P*-value found in each window and then compare the average 10%, 5%, 2% and 1% lowest *P*-value windows near and far from amino acid substitutions. * randomization test $P \leq 0.05$; ** $P \leq 0.01$; *** $P \leq 0.001$. Left side of histograms: all regions irrespective of recombination rates. Rigth side: only regions with recombination rates lower than 0.5 cM/Mb. The max(EA,EU) histograms show the results obtained when retaining for each window the maximum signal of the East Asian and European populations.

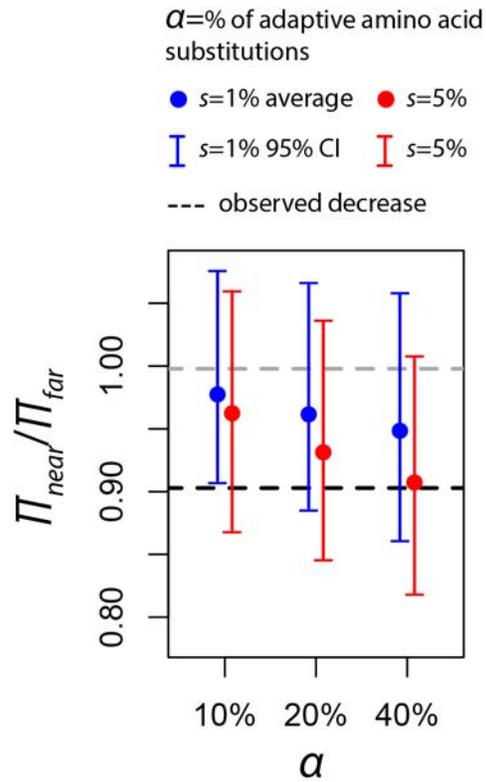

**Figure 6. Simulated decreases of diversity for different rates and strengths of positive selection**
We ran 100 forward simulations (Methods) to estimate the average and 95% confidence intervals (CI) for the decrease of diversity near amino acid changes under different rates and strengths of positive selection. To be conservative, we extended the confidence intervals from simulations by 35% given that neutral simulations underestimate the variance by ~35% in the simulated regions as shown in Fig. 1.

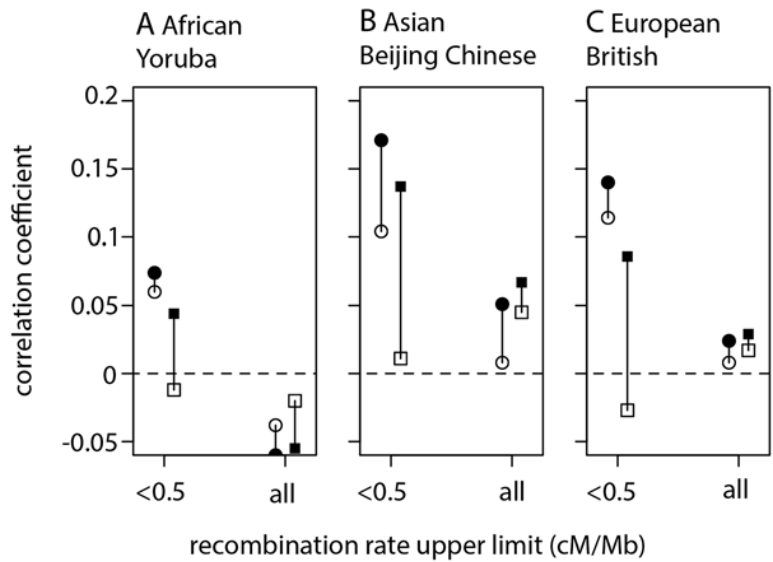

**Figure 7. Most human recent positive selection occurs in regulatory sequences**

The filled circles and squares show the correlation coefficients of the absolute values of *iHS* with the density of regulatory and coding sequence density, respectively, controlling for recombination and average pairwise diversity (Methods). The open circles and squares show partial correlations. For instance, an open circle shows partial correlation between absolute values of *iHS* and regulatory density controlling for coding density (also controlling for recombination and average pairwise diversity). All correlations are Spearman's rank correlations or partial correlations. Correlation coefficients greater than 0.05 are all highly significant ($P<2\times10^{-16}$).

**References**

Abecasis GR, Auton A, Brooks LD, DePristo MA, Durbin RM, Handsaker RE, Kang HM, Marth GT, McVean GA. 2012. An integrated map of genetic variation from 1,092 human genomes. *Nature* **491**(7422): 56-65.

Adzhubei IA, Schmidt S, Peshkin L, Ramensky VE, Gerasimova A, Bork P, Kondrashov AS, Sunyaev SR. 2010. A method and server for predicting damaging missense mutations. *Nat Methods* **7**(4): 248-249.

Akey JM. 2009. Constructing genomic maps of positive selection in humans: where do we go from here? *Genome Res* **19**(5): 711-722.

Andolfatto P. 2007. Hitchhiking effects of recurrent beneficial amino acid substitutions in the Drosophila melanogaster genome. *Genome Res* **17**(12): 1755-1762.

Boyko AR, Williamson SH, Indap AR, Degenhardt JD, Hernandez RD, Lohmueller KE, Adams MD, Schmidt S, Sninsky JJ, Sunyaev SR et al. 2008. Assessing the evolutionary impact of amino acid mutations in the human genome. *PLoS Genet* **4**(5): e1000083.

Cai JJ, Macpherson JM, Sella G, Petrov DA. 2009. Pervasive hitchhiking at coding and regulatory sites in humans. *PLoS Genet* **5**(1): e1000336.

Charlesworth B, Morgan MT, Charlesworth D. 1993. The effect of deleterious mutations on neutral molecular variation. *Genetics* **134**(4): 1289-1303.

Eyre-Walker A, Keightley PD. 2009. Estimating the rate of adaptive molecular evolution in the presence of slightly deleterious mutations and population size change. *Mol Biol Evol* **26**(9): 2097-2108.

Fletcher W, Yang Z. 2010. The effect of insertions, deletions, and alignment errors on the branch-site test of positive selection. *Mol Biol Evol* **27**(10): 2257-2267.

Gerstein MB, Kundaje A, Hariharan M, Landt SG, Yan KK, Cheng C, Mu XJ, Khurana E, Rozowsky J, Alexander R et al. 2012. Architecture of the human regulatory network derived from ENCODE data. *Nature* **489**(7414): 91-100.

Grossman SR, Andersen KG, Shlyakhter I, Tabrizi S, Winnicki S, Yen A, Park DJ, Griesemer D, Karlsson EK, Wong SH et al. 2013. Identifying recent adaptations in large-scale genomic data. *Cell* **152**(4): 703-713.

Grossman SR, Shlyakhter I, Karlsson EK, Byrne EH, Morales S, Frieden G, Hostetter E, Angelino E, Garber M, Zuk O et al. 2010. A composite of multiple signals distinguishes causal variants in regions of positive selection. *Science* **327**(5967): 883-886.

Hernandez RD, Kelley JL, Elyashiv E, Melton SC, Auton A, McVean G, Sella G, Przeworski M. 2011. Classic selective sweeps were rare in recent human evolution. *Science* **331**(6019): 920-924.

Keightley PD, Eyre-Walker A. 2007. Joint inference of the distribution of fitness effects of deleterious mutations and population demography based on nucleotide polymorphism frequencies. *Genetics* **177**(4): 2251-2261.

Kent WJ. 2002. BLAT--the BLAST-like alignment tool. *Genome Res* **12**(4): 656-664.

King MC, Wilson AC. 1975. Evolution at two levels in humans and chimpanzees. *Science* **188**(4184): 107-116.

Kong A, Thorleifsson G, Gudbjartsson DF, Masson G, Sigurdsson A, Jonasdottir A, Walters GB, Gylfason A, Kristinsson KT, Gudjonsson SA et al. 2010. Fine-scale recombination rate differences between sexes, populations and individuals. *Nature* **467**(7319): 1099-1103.

Lander ES Linton LM Birren B Nusbaum C Zody MC Baldwin J Devon K Dewar K Doyle M FitzHugh W et al. 2001. Initial sequencing and analysis of the human genome. *Nature* **409**(6822): 860-921.

Lohmueller KE, Albrechtsen A, Li Y, Kim SY, Korneliussen T, Vinckenbosch N, Tian G, Huerta-Sanchez E, Feder AF, Grarup N et al. 2011. Natural selection affects multiple aspects of genetic variation at putatively neutral sites across the human genome. *PLoS Genet* **7**(10): e1002326.

Loytynoja A, Goldman N. 2008. Phylogeny-aware gap placement prevents errors in sequence alignment and evolutionary analysis. *Science* **320**(5883): 1632-1635.

Macpherson JM, Sella G, Davis JC, Petrov DA. 2007. Genomewide spatial correspondence between nonsynonymous divergence and neutral polymorphism reveals extensive adaptation in Drosophila. *Genetics* **177**(4): 2083-2099.

McVicker G, Gordon D, Davis C, Green P. 2009. Widespread genomic signatures of natural selection in hominid evolution. *PLoS Genet* **5**(5): e1000471.

Messer PW. 2013. SLiM: simulating evolution with selection and linkage. *Genetics* `(in press: published as early access 0.1534/genetics.113.152181).` .

Messer PW, Petrov DA. 2013. Frequent adaptation and the McDonald-Kreitman test. *Proc Natl Acad Sci U S A* **110**(21): 8615-8620.

Pickrell JK, Coop G, Novembre J, Kudaravalli S, Li JZ, Absher D, Srinivasan BS, Barsh GS, Myers RM, Feldman MW et al. 2009. Signals of recent positive selection in a worldwide sample of human populations. *Genome Res* **19**(5): 826-837.

Przeworski M. 2002. The signature of positive selection at randomly chosen loci. *Genetics* **160**(3): 1179-1189.

Sabeti PC, Schaffner SF, Fry B, Lohmueller J, Varilly P, Shamovsky O, Palma A, Mikkelsen TS, Altshuler D, Lander ES. 2006. Positive natural selection in the human lineage. *Science* **312**(5780): 1614-1620.

Sabeti PC Varilly P Fry B Lohmueller J Hostetter E Cotsapas C Xie X Byrne EH McCarroll SA Gaudet R et al. 2007. Genome-wide detection and characterization of positive selection in human populations. *Nature* **449**(7164): 913-918.

Sattath S, Elyashiv E, Kolodny O, Rinott Y, Sella G. 2011. Pervasive adaptive protein evolution apparent in diversity patterns around amino acid substitutions in Drosophila simulans. *PLoS Genet* **7**(2): e1001302.

Schwartz S, Kent WJ, Smit A, Zhang Z, Baertsch R, Hardison RC, Haussler D, Miller W. 2003. Human-mouse alignments with BLASTZ. *Genome Res* **13**(1): 103-107.

Siepel A, Bejerano G, Pedersen JS, Hinrichs AS, Hou M, Rosenbloom K, Clawson H, Spieth J, Hillier LW, Richards S et al. 2005. Evolutionarily conserved elements in vertebrate, insect, worm, and yeast genomes. *Genome Res* **15**(8): 1034-1050.

Voight BF, Kudaravalli S, Wen X, Pritchard JK. 2006. A map of recent positive selection in the human genome. *PLoS Biol* **4**(3): e72.

Williamson SH, Hubisz MJ, Clark AG, Payseur BA, Bustamante CD, Nielsen R. 2007. Localizing recent adaptive evolution in the human genome. *PLoS Genet* **3**(6): e90.

**Supplemental Material**

**The omnibus near versus far and functional versus non-functional tests reveals a strong departure from neutrality**

The near versus far test (Fig. 1) and the functional versus non-functional tests (Fig. 2) look for different signals in the data and should be independent of one another. Indeed, all regions in the functional versus non-functional test are located near an amino acid substitution and thus they all are in the "near" category in the near versus far test. The fact that the near regions have lower diversity than the far regions should not affect the results of the test that looks within the near regions. In addition, we confirmed that the finite number of regions used in the bootstrap procedure does not generate spurious correlations between the two tests. To ensure that the near versus far and functional versus non-functional tests are independent, we verify that their results are uncorrelated. To do so we run 500 neutral population simulations using SliM (rescaling factor 20, see Methods). From these 500 simulations we calculate (as for Figs. 1 and 2) 500 pairs of $\pi_{near}/\pi_{far}$ and $\pi_{func}/\pi_{non-func}$ and calculate the correlation between the two values. As expected there is no correlation between the two ratios ($n$=500, Spearman's $\rho$=5x10$^{-4}$ $P$=0.99), which shows that the two tests are completely independent. The independence of the two tests enables us to run the randomization test separately for the *near/far* and for the *functional/non-functional* tests and combine the results. To do so we run 10,000 iterations of our randomization procedure separately for both tests and we couple *near/far* iteration *n* with f*unctional/non-functional* iteration *n*. This is made possible by the fact that the two tests are independent. In all human populations together the observed combined decreases are highly unexpected, as shown by the $P$-values of the combined randomization test on Fig. 3. Even in Africa where the signal of positive selection is consistently weaker in both the near versus far and the functional versus non-functional tests, the probability of both observed decreases by chance is less than 1%. In European and Asian populations the same probability is lower than 0.1%. Taken together these results strongly suggest that positive selection has significantly decreased neutral diversity in the human genome.

**Testing the robustness of *iHS* and *XPEHH* to BGS**

The strong sensitivity of average heterozygosity to BGS makes it very challenging to distinguish the specific effects of positive selection using this particular measure of diversity. Indeed, we could only detect a signature of selective sweeps when BGS is weak, that is in regions with a low density of conserved coding sequences (Results). Measures of diversity such as *iHS* and *XPEHH* should be much more robust to BGS than average heterozygosity. Indeed they are based on the frequency and length of haplotypes, and there is no *a priori* reason why BGS should affect haplotype length and frequency. In particular, BGS is not expected to create the long and frequent haplotypes that *iHS* and *XPEHH* are sensitive to.

In order to test the robustness of *iHS* and *XPEHH* to BGS, we use Slim to simulate 4 Mb sequences that include a 100 kb central region where deleterious mutations occur with a predefined strength of selection and rate. We simulate 4 Mb which is large to avoid edge effects in the calculations of *iHS* and *XPEHH*. For this reason we exclude the first and last megabases of the simulated sequences. The population size is 1,000, the rescaled mutation rate is set at $10^{-7}$ and the rescaled recombination rate is uniform and set at 10 cM/Mb.

We test the effect of different BGS configurations on average heterozygosity, *iHS* and *XPEHH*. First, we test the effect of having 25% of the mutations in the 100kb central region with *Nes*=-200, which corresponds to strongly deleterious mutations (Supplemental Fig.1 first column). The same is done this time with 50% of strongly deleterious mutations (Supplemental Fig. 1 second column). We then test the effect of having 25% of weakly deleterious mutations with *Nes*=-5 (third column). The same is repeated with 50% of weakly deleterious mutations (fourth column). Finally we test the effect of deleterious mutations with gamma-distributed intensities of selection. Parameters of the gamma distribution fit current estimates of the distribution of fitness effects in human (Keightley and Eyre-Walker 2007). In this case we test 25% (fifth column) of mutations being deleterious, which results in a 20% decrease of heterozygosity that matches the average decrease observed near coding sequences in human (McVicker et al. 2009). For each condition we run 500 independent simulations to obtain averages and confidence intervals.

    In every case tested, we find that BGS has virtually no effect on *XPEHH*, and a weak, conservative effect on *iHS* (Supplemental Fig. 1).

**Correlations between *iHS*, ENCODE regulatory elements and coding sequences**

We investigate the correlations between *iHS* (Voight et al. 2006) and ENCODE

regulatory elements (ERE) and between *iHS* and coding sequences to test whether regulatory elements are the main source of adaptive mutations in the human genome. ERE density in our analysis is the density of elements predicted as DNASE1 hypersensitive sites and also as transcription factor binding sites identified by Chip-Seq by the ENCODE Consortium (Gerstein et al. 2012). Using the overlap of both prediction methods ensures that we work with higher confidence ERE. In this analysis we calculate *iHS* using the latest 1,000 Genomes project data for three human population (Yoruba, British and Beijing Chinese phase 1 July 2012 release) instead of using publicly available results from published scans. Indeed in this correlation analysis we need to calculate the correlation between values of *iHS* and values of average heterozygosity that have to be calculated from the same set of variants. To calculate this correlation we use 1 Mb windows sliding every 50 kb along the human genome. For this analysis we use 1 Mb instead of 500 kb windows because the correlation between coding density and *iHS* is slightly higher when using 1 Mb windows (windows of recombination rate lower than 0.5 cM/Mb; 1 Mb $n$=9,471; Yoruba Spearman's $\rho$=0.044 $P$<$2\times10^{-16}$; British $\rho$=0.086 $P$<$2\times10^{-16}$; Beijing Chinese $\rho$=0.137 $P$<$2\times10^{-16}$; 500 kb $n$=12,515; Yoruba Spearman's $\rho$=0.012 $P$=0.14; British $\rho$=0.072 $P$<$2\times10^{-16}$; Beijing Chinese $\rho$=0.107 $P$<$2\times10^{-16}$). It is important that we use the window size that maximizes the correlation between *iHS* and coding density since we later evaluate the effect of controlling for coding density on the correlation between *iHS* and ERE. In addition windows less than 5 Mb from centromeres and telomeres or with more than 20% assembly gaps are removed from the correlation analysis.

      In each window positive selection may result in extreme *iHS* values only for a small subset of variants. Thus for each window we select only the top 5% absolute *iHS* variants and measure the average absolute *iHS* of this top 5%. This is done separately for 50 individuals in the British population (GBR, first 50 individual in the July 2012 phase 1 VCF files), 50 individuals in the Chinese Beijing population (CHB, first 50 individual in the July 2012 phase 1 VCF files) and 50 individuals in the Yoruba population (YRI, first 50 individual in the July 2012 phase 1 VCF files). We use a limited number of individuals in each population so that the calculations of *iHS* are computationally feasible in a reasonable amount of time.

      We first calculate the correlation between *iHS* (top 5% average) and average heterozygosity to confirm the robustness of *iHS* to BGS previously deduced from our forward simulations. In all the three populations, the top 5% average *iHS* correlates

positively with average heterozygosity (windows with recombination rate lower than 0.5 cM/Mb; $n$=9,471; YRI $\rho$=0.073 $P$<2x10$^{-16}$; CHB $\rho$=0.117 $P$<2x10$^{-16}$; GBR $\rho$=0.144 $P$<2x10$^{-16}$). Thus *iHS* is conservative relative to BGS, as already observed in population simulations. We correct for this conservativeness by controlling for average heterozygosity when calculating the correlations between *iHS* and coding density and between *iHS* and ERE density (see above and Fig. 7). Therefore all the correlations in the analysis are calculated controlling for recombination rate and average heterozygosity, except for the correlation between *iHS* and average heterozygosity, where we control only for recombination rate.

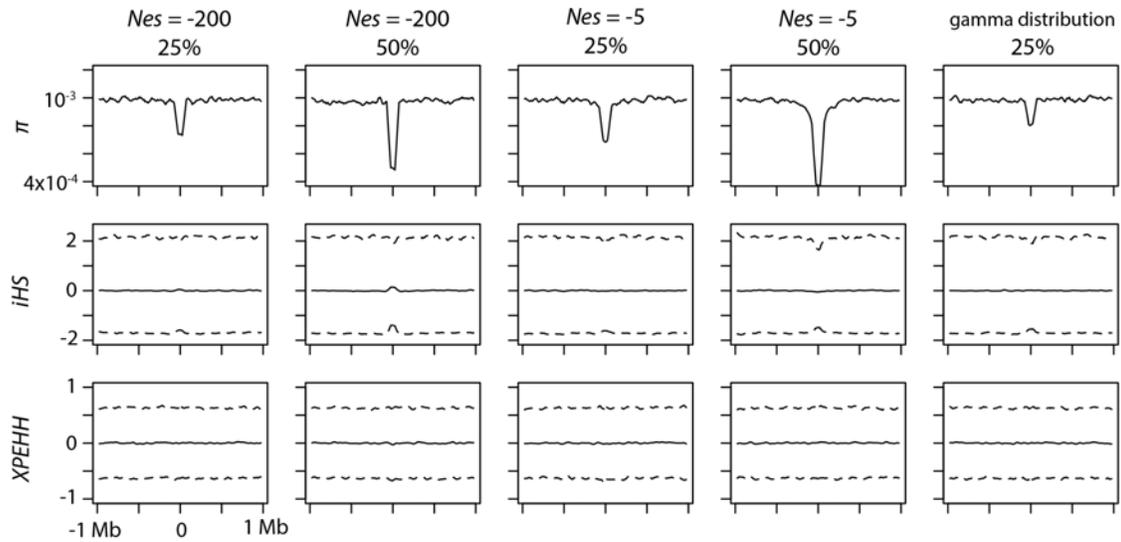

**Supplemental Figure 1. Robustness of *iHS* and *XPEHH* to various rates and strengths of BGS**

We tested the effect of BGS on *iHS* and *XPEHH* (Results and Supplemental Material). Upper row: average heterozygosity. Middle row: *iHS*. Lower row: *XPEHH*. The full line represents the average *iHS* or *XPEHH* along the simulated region. The dashed lines represent the limits of the *iHS* or *XPEHH* 95% confidence intervals.

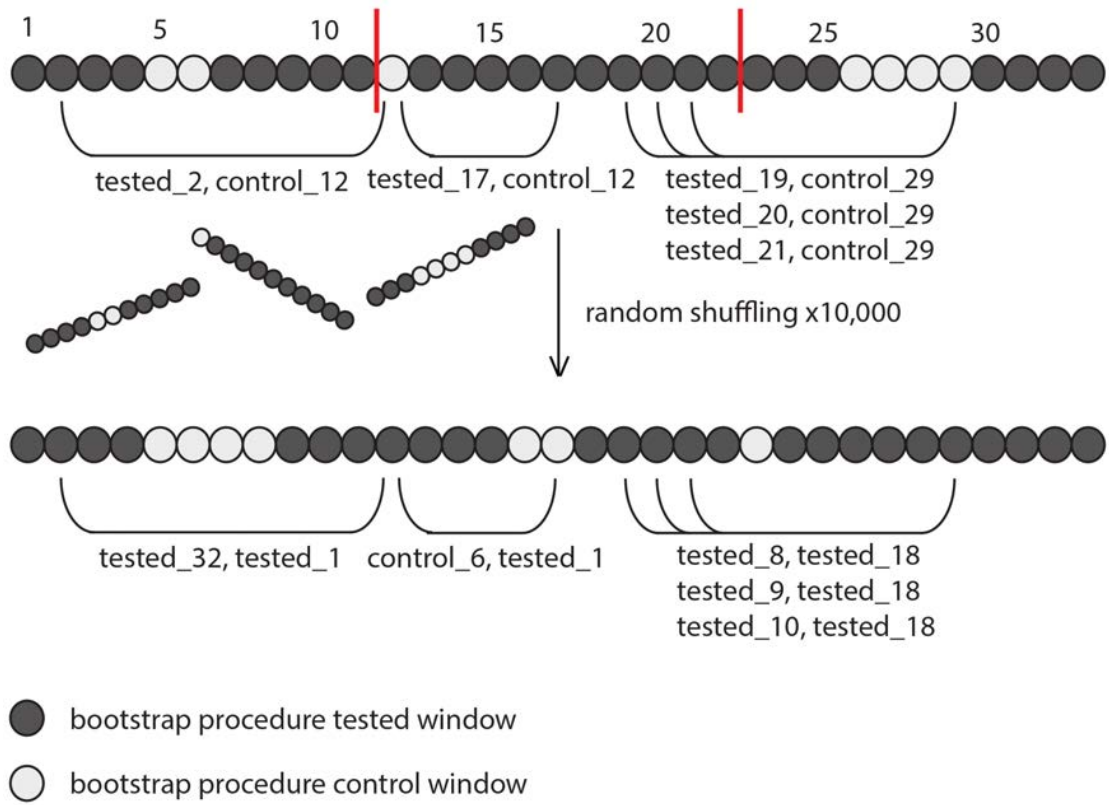

**Supplemental Figure 2. Randomization test scheme.**

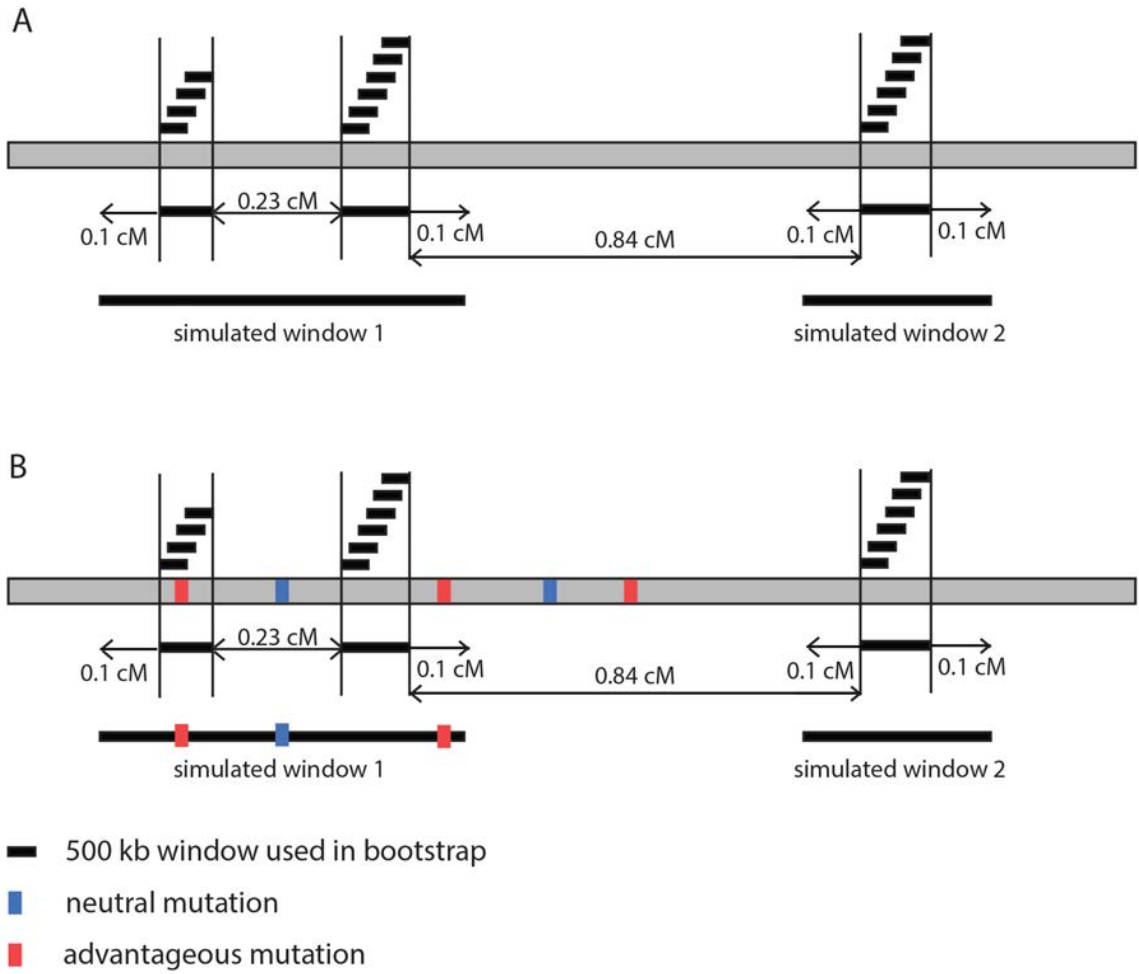

**Supplemental Figure 3. Population simulation scheme.**

**Supplemental Table 1. Amount of sequences used in the bootstrap procedure**[*]

| Recombination rate | <0.5 cM/Mb | <1 cM/Mb | <1.5 cM/Mb | <2 cM/Mb | all |
|---|---|---|---|---|---|
| Near vs far test | na | 614-105 | 838-185 | 971-262 | 1149-463 |
| near vs far test CCDS<0.5% | na | 136-57 | 204-105 | 242-149 | 290-236 |
| functional vs non-functional test | na | 523-320 | 662-471 | 733-587 | 823-768 |
| functional vs non-functional test CCDS<0.5% | na | 103-101 | 129-134 | 141-150 | 157-179 |
| near vs far test *XPEHH iHS CLR* | 665-87 | na | na | na | 1562-619 |

[*]This table provides the amount of sequence used in the different tests conducted. In the grey-shaded cell 614-105 means the following: for the near vs test using only windows with a recombination rate lower than 1 cM/Mb, 614 megabases of sequences near amino acid changes are used, and 105 megabases of sequences far from amino acid changes are used.

**Supplemental Table 2. Thresholds used for tested/control window matching in the bootstrap procedure**[*]

|  | near vs far | functional vs non-functional | near vs far *XPEHH iHS CLR* |
|---|---|---|---|
| Recombination rate | 50%-150% | 50%-150% | 10%-160% |
| GC content | 97.2%-1.028% | 95.3%-105% |  |
| CDS | 90%-∞ | 75%-1.45% | 70%-500% |
| CCDS | 10%-125% | 70%-200% |  |
| surrounding CDS | 40%-∞ | 45%-∞ |  |
| UTR | 70%-160% | 70%-160% |  |
| TFD | 80%-130% | 65%-140% |  |

[*]This table provides the thresholds used for choosing control windows as a function of the values of different variables of the tested windows (first column). The grey-shaded cell means that in the near versus far test recombination rate within a control window has to be comprised between 50% and 150% of the recombination rate within the tested window.